\documentstyle[12pt]{article}
\newcommand{\be}{\begin{equation}}
\newcommand{\ee}{\end{equation}}
\newcommand{\bea}{\begin{eqnarray}}
\newcommand{\eea}{\end{eqnarray}}
\def\pd{\partial}
\makeatletter
\def\eqnarray{\stepcounter{equation}\let\@currentlabel=\theequation
\global\@eqnswtrue
\global\@eqcnt\z@\tabskip\@centering\let\\=\@eqncr
$$\halign to \displaywidth\bgroup\@eqnsel\hskip\@centering
  $\displaystyle\tabskip\z@{##}$&\global\@eqcnt\@ne
  \hfil$\displaystyle{\hbox{}##\hbox{}}$\hfil
  &\global\@eqcnt\tw@ $\displaystyle\tabskip\z@
  {##}$\hfil\tabskip\@centering&\llap{##}\tabskip\z@\cr}
\@addtoreset{equation}{section}
  \def\theequation{\thesection.\arabic{equation}}
\makeatother

\begin{document}

\def\pd{\partial}
\def\ts{\thinspace}
\vskip 20pt
\centerline{\bf A MODEL FOR CLASSICAL SPACE-TIME CO-ORDINATES}
\vskip  20pt
\centerline{\bf{David B.Fairlie}\footnote{Currently on leave of absence at {\it
TH division, CERN, Geneva}}\footnote{e-mail david.fairlie@durham.ac.uk}}
\vskip 0.5 true cm
\centerline{Department of Mathematical Sciences,}
\centerline{University of Durham, Durham, DH1 3LE, England.}
\vskip 20pt
\centerline{\bf{Renat Zhdanov}\footnote{On leave of absence from
{\it Institute of Mathematics, Tereshchenkivska Street 3, 252004 Kiev, Ukraine.}}
\footnote{e-mail rzhdanov@apmat.freenet.kiev.ua}}
\vskip 0.5 true cm
\centerline{ AS Institute for Mathematical Physics,}
\centerline{TU Clausthal, Leibnizstrasse  10,}
\centerline{ 38678  Clausthal-Zellerfeld, Germany.}
\vskip 2 true cm
\noindent Abstract :
Field equations with general covariance are interpreted as equations for
a target space describing physical space time co-ordinates, in terms of an
underlying base space with conformal invariance. These equations admit an
infinite number of inequivalent Lagrangian descriptions. A model for
reparametrisation invariant membranes is obtained by reversing
the roles of base and target space variables in these considerations.

\vfill\eject

\section{Introduction.}
\smallskip
A characteristic feature of the classical equations of General Relativity is the
property of General Covariance; i.e that the equations are covariant under
differentiable re-definitions of the space-time co-ordinates. In  the first of a
series of papers investigating a class of covariant equations which
Jan Govaerts and the first author, which we called `Universal Field Equations'
~{\cite{fai}--\cite{5}} we floated the idea that these equations could be employed
as a model for space time co-ordinates. It is one object of this paper to explore
this idea in somewhat greater depth. This is a purely classical discussion of
a way of describing a co-ordinate system which is sufficiently flexible to
admit the general class of functional redefinitions implied by covariance.
It has nothing to do with quantum effects
like the concept of a minimum compactification radius due to T duality which
rules out the the notion of an infinitely precise point in space time. Here the
discussion will remain entirely classical and will explore the idea that the
space-time co-ordinates in $D$ dimensions may be represented by `flat' co-ordinates
in $D+1$ dimensions, which transform under the conformal group in
$D+1$ dimensions. There are, however two ways to implement general covariance;
one by the use of covariant derivatives, and the other by exploting properties of
determinants. In a second application the `Universal Field Equations'
may be regarded as describing membranes, by reversing the roles of fields and
base-co-ordinates. Then the covariance of fields becomes the reparametrisation invariance
of the new base space.
\section{Multifield UFE}
Suppose $X(x_i)^a,\ a=1,\dots ,D ,\ i=1,\dots ,D+1 $ denotes a set of $D$ fields,
in $D+1$ dimensional space. They may be thought of as  target space co-ordinates
which constitute a mapping from a $D+1$ dimensional base space co-odinatized by the
independent variables $x_i$. Introduce the notation $\displaystyle{X^a_i=\frac{\pd X^a}
{\pd x_i},\ X^a_{ij}=\frac{\pd^2 X^a}{\pd x_i\pd x_j}}$. In addition, let
$J_k$ denote the Jacobian
$\displaystyle{\frac{\pd (X^a,X^b,\dots,X^D)}{\pd(x_1,\dots,\hat x_k\dots ,x_{D+1})}}$
where $x_k$ is the independent variable which is omitted in $J_k$.
Now suppose that the vector field $X^a$ satisfies the equations of motion
\be
\sum_{i,k}J_iJ_kX^a _{ik}=0.
\label{eqmo}
\ee
This is a direct generalisation of the Bateman equation to $D$ fields in $D+1$
dimensions, \cite{fai}, and may be written in terms of the determinant of a bordered
matrix where the diagonal blocks are of dimensions $D\times D$ and $D+1\times D+1$
respectively as
\be
\det\left\|\begin{array}{cc} 0&\frac{\pd X^a}{\pd x_k}\\
  \frac{\pd X^b}{\pd x_j}&\sum\lambda_c\frac{\pd^2 X^c}{\pd x_j\pd x_k}\end{array}\right\|=0.
\label{deqmo}
\ee
The coefficients of the arbitrary constant parameters $\lambda_c$ set to
zero reproduce the $D$ equations (\ref{eqmo}). The solutions of these equations
can be verified to possess the property that any functional redefinition of
a specific solution is also a solution; i.e. the property of general covariance.
A remarkable feature of  (\ref{eqmo}) is that the equations admit infinitely many
inequivalent Lagrangian formulations. Suppose ${\cal L}$ depends upon
the fields $X^a$ and their first derivatives $X^a_j$ through the Jacobians  subject
only to the constraint that  ${\cal L}(X^a,\ J_j)$  is a homogeneous function of the
Jacobians, i.e.
\be
\sum_{j=1}^{D+1}J_j\frac{\pd {\cal L}}{\pd J_j}={\cal L}.
\label{hom}
\ee
Then the Euler variation of ${\cal L}$ with respect to the field $X^a$ gives
\bea \label{euler}
& &\frac{\pd{\cal L}}{\pd X^a}-\frac{\pd}{\pd x_i}\frac{\pd{\cal L}}{\pd X^a_i}\nonumber\\
&=&
\frac{\pd{\cal L}}{\pd X^a}-\frac{\pd}{\pd x_i}\frac{\pd{\cal L}}{\pd J_j}\frac{\pd J_j}
{\pd X^a_i}\\
&=&\frac{\pd{\cal L}}{\pd X^a}-\frac{\pd^2{\cal L}}{\pd X^b\pd J_j}\frac{\pd J_j}
{\pd X^a_i}X^b_i-\frac{\pd{\cal L}}{\pd J_j}\frac{\pd^2 J_j}{\pd X^a_i\pd X^b_k}
X^b_{ik}-\frac{\pd^2{\cal L}}{\pd J_j\pd J_k}\frac{\pd J_j}{\pd X^a_i}\frac{\pd J_k}
{\pd X^b_r}X^b_{ir}.\nonumber
\eea
The usual convention of summing over repeated indices is adhered to here.
Now by the theorem of false cofactors
\be
\sum_{j=1}^{D+1}\frac{\pd J_k}{\pd X^a_j}X^b_j = \delta_{ab}J_k.
\label{falseco}
\ee
Then, exploiting the homogeneity of ${\cal L}$ as a function of $J_k$ (\ref{hom}),
the first two terms in the last line of (\ref{euler}) cancel, and the term
$\displaystyle{\frac{\pd{\cal L}}{\pd J_j}\frac{\pd^2 J_j}{\pd X^a_i\pd X^b_k}
X^b_{ik}}$ vanishes by symmetry considerations. The remaining term,
$\displaystyle{\frac{\pd^2{\cal L}}{\pd J_j\pd J_k}\frac{\pd J_j}{\pd X^a_i}
\frac{\pd J_k}{\pd X^b_r}X^b_{ir}}$, may be simplified as follows.
Differentiation of the homogeneity equation (\ref{hom}) gives
\be\sum^{D+1}_{k=1} \frac{\pd^2{\cal L}}{\pd J_j\pd J_k}J_k = 0.
\label{hom1}
\ee
But since $\sum_k J_kX^a_k=0,\ \forall a$,  together with symmetry,
this implies that the linear equations (\ref{hom1}) can be solved by
\be
\frac{\pd^2{\cal L}}{\pd J_i\pd J_j}= \sum_{a,b}X^a_i d^{ab}X^b_j,
\label{hom2}
\ee
for some  functions $d^{ab}$. Inserting this representation into (\ref{euler}) and
using a similar result to (\ref{falseco});
\be
\sum_{j=1}^{D+1}\frac{\pd J_j}{\pd X^a_k}X^b_j = -\delta_{ab}J_k.
\label{false}
\ee
Then, assuming $d^{a,b}$ is invertible, as is the
generic case, the last term reduces to $\sum_{i,k}J_iJ_kX^a _{ik}$,
which, set to zero is just the equation of motion (\ref{eqmo})\footnote
{This calculation  without the $X^a$ dependence of the Lagrangian  already can be
found in \cite{fai}; the new aspect here is the extension to include the fields
themselves, following the single field example of \cite{jam}.}
\subsection{Iteration}
This procedure may be iterated; Given a transformation described by the equation
(\ref{eqmo}), from a base space of $D+2$ dimensions with co-ordinates $x_i$ to to a
target space of $D+1$ with co-ordinates $Y_j$ which in turn are used as a
base space for a similar transformation to co-ordinates $X_k,\ k=1\dots D$ the
mapping from $D+1$ dimensions to $D$ is given in terms of  the determinant of a
bordered matrix of similar form to (\ref{deqmo}), where the diagonal blocks are
of dimensions $D\times D$ and $D+2\times D+2$ respectively;
\be
\det\left\|\begin{array}{cc} 0&\frac{\pd X^a}{\pd x_k}\\
  \frac{\pd X^b}{\pd x_j}&\sum\lambda_j\frac{\pd^2 X^j}{\pd x_j\pd x_k}
\end{array}\right\|=0.
\label{feqmo}
\ee
The equations which form an overdetermined set are obtained by requiring that
the determinant vanishes for all choices of $\lambda_j$
Further iterations yield the multifield UFE, discussed in \cite{4}, and the
Lagrangian description is given by a iterative procedure.
\subsection{\bf Solutions.}
There are various ways to approach the question of solutions. Consider
the multifield UFE;
\begin{equation}
\label{1}
{\rm det}\,\left\|
\begin{array}{cccccc}
0&\ldots&0&X^1_{x_1}&\ldots&X^1_{x_d}\\
\vdots&\ddots&\vdots&\vdots&\ddots&\vdots\\
0&\ldots&0&X^n_{x_1}&\ldots&X^n_{x_d}\\
X^1_{x_1}&\ldots&X^n_{x_1}&\sum_{i=1}^n\lambda_iX^i_{x_1x_1}&\ldots&\sum_{i=1}^n
\lambda_iX^i_{x_1x_d}\\
\vdots&\ddots&\vdots&\vdots&\ddots&\vdots\\
X_{x_d}&\ldots&X_{x_d}&\sum_{i=1}^n\lambda_iX^i_{x_1x_d}&\ldots&\sum_{i=1}^n
\lambda_iX^i_{x_dx_d}
\end{array}\right \|=0,
\ee
where $\lambda_1,\dots,\lambda_n$ are arbitrary constants, and the functions
$X^1,\dots,X^n$ are independent of $\lambda_i$. The equations which result
from setting the coefficients of the monomials of degree $d-n$ in $\lambda_i$
in the expansion of the determinant to zero form an overdetermined set, but,
as we shall show, this set possesses many nontrivial solutions.

\noindent
The equation (\ref{1}) may be viewed as a special case of the Monge-Amp\`ere
equation in $d+n$ dimensions, namely
\begin{equation}
\label{2}
  {\rm det}\, \left \|{\partial^2 u\over
      \partial_{y_i}\partial_{y_j}}\right \|^{d+n}_{i,j=1}=0 .
\end{equation}
Equation (\ref{1}) results from the restriction of $u$ to have the form
\be
u(y_k)=u(x_1,\dots,x_d,\lambda_1,\dots,\lambda_n)=\sum_{i=1}^n\lambda_iX^i,
\label{three}
\ee
where we have set
\be
y_i=x_i,\ i=1,\dots,d,\ \ y_{j+d}=\lambda_{j},\ j=1,\dots,n.
\label{monge}
\ee
Now the Monge-Amp\`ere equation is equivalent to the statement that there
exists a functional dependence among the first derivatives $u_{y_i}$ of $u$ of the form
\be
F(u_{y_1},\dots,u_{y_{d+n}})=0,
\label{monge2}
\ee
where $F$ is an arbitrary differentiable function. Methods for the solution of
this equation are known~\cite{ren,dbf}.
Returning to the target space variables $X^j$, this relation becomes
\begin{equation}
\label{4}
F\left(\underbrace{\sum_{i=1}^n\lambda_iX^i_{x_1}}_{\omega_1},\dots,
\underbrace{\sum_{i=1}^n\lambda_iX^i_{x_d}}_{\omega_d},\ X^1,\ldots,X^n\right)=0.
\end{equation}
\section{Exact Solutions of the UFE}
\subsection{Implicit Solutions}
The general representation of a solution of this set of constraints which do
not depend upon the parameters $\lambda^i$ evades us; however there are two
circumstances in which a solution may be found. In the first case a class of
solutions in implicit form may be obtained by taking $F$ to be linear in the
first $d$ arguments $\omega_i$.
Then
\be
F=\sum^d_{i=1}f_i(X^1,\ldots,X^n)\omega_i=0.
\label{linear}
\ee
It can be proved that this is the generic situation for the cases of two and
three fields. In general, provided there are terms linear in $\lambda_i$ in
$F$, as the $X^i$ do not depend
upon $\lambda_i$, one expects that as a minimal requirement the
terms in $F$ linear in $\lambda_i$ will vanish for a solution.
Equating each  coefficient of $\lambda^i$ in (\ref{linear}) to zero we obtain
the following system of partial differential equations
\be
\sum^d_{i=1}f_i(X^1,\ldots,X^n)X^j_{x_i}=0,\ \ j=1,\dots,n.
\label{system}
\ee
The general solution of these equations may be represented in terms of $n$ arbitrary
smooth functions $R^j$, where
\be
R^j(f_dx_1-f_1x_d,\dots,f_dx_{d-1}-f_{d-1}x_d,\ X^1,\dots,X^n)=0.
\ee
The solution of these equations  for $X^i$ gives a wide class of solutions to
the UFE.
\subsection{Explicit Solution.}
There is a wide class of explicit solutions to the UFE. They are simply given by
choosing $X^j(x_1,\dots,x_d)$ to be a homogeneous function of $x_j$ of weight zero, i.e.
\be
\sum_{k=1}^dx_k\frac{\pd X^j}{\pd x_k}=0,\ \ j=1,\dots,n.
\label{explicit}
\ee
The proof of this result depends upon differentiation of (\ref{explicit}) with respect
to the $x_i$.
A particularly illuminating example is the case of spherical polars;
in $d=3,\ n=2$ take
\be
X^1=\phi =\arctan\left(\frac{x_3}{\sqrt{x_1^2+x_2^2}}\right) ;\
X^2=\theta =\arctan\left(\frac{x_2}{x_1}\right).
\label{sphere}
\ee
Then these co-ordinates satisfy (\ref{feqmo}).

\section{Conclusions}
A wide class of solutions to the set of UFE which are generally covariant
has been obtained. In order to adapt the theory to apply to possible
integrable membranes, it is necessary to interchange the roles of dependent
and independent variables, so that general covariance becomes reparametrisation
invariance of the base space~\cite{3}. In order to invert the dependent and
independent variables in this fashion, it is necessary  first to augment the
dependent variables by some additional $d-n$ fields $Y_k(x_i)$, then consider
the $x_i$ as functions of $X_j,\ i=1\dots n$ Although in principle $x_i$ could
also depend upon the artificial variables $Y_k,\ k=1\dots d-n$, we make the
restriction that this does not occur. (See \cite{3} for further details)
 In this case the
variables $x_j$ play the role of target space for an $n$-brane, dependent upon
$n$ co-ordinates $X^j$. Since it is fully reparametrisation invariant, it may
play some part in the further understanding of string theory, but
this is by no means clear.
\section{Acknowledgement}
Renat Zhdanov would like to thank the "Alexander von Humboldt Stiftung"
for financial support.

\end{document}